\documentstyle[epsf]{npbproc}
% ...
\newcommand{\RU}{Department of Physics and Astronomy, Rutgers
            University, Piscataway, NJ 08855-0849}
\newcommand{\SCRI}{SCRI, The Florida State University,
                  Tallahassee, FL 32306, USA.}

\begin{document}

\title{HOW TO PUT A HEAVIER HIGGS ON THE LATTICE \thanks%
{Presented by P. ~Vranas}}
\author{ U.~M.~Heller${}^1$,
         M.~Klomfass${}^1$, 
         H.~Neuberger${}^{2}$,
         P.~Vranas${}^1$}
\address{${}^1$\SCRI\\
         ${}^3$\RU}

\date{}

\runtitle{A heavier Higgs on the lattice} 
\runauthor{U. M. Heller et al.} 
\volume{XXX}  
\firstpage{1} 
\lastpage{3}  

\begin{abstract}

The cutoff dependence of the Scalar Sector of the Minimal Standard
Model can result in an increase of the existing triviality bound
estimates of the Higgs mass. We present a large $N$ calculation and
some preliminary $N=4$ results that suggest that the increase can be
as large as $30\%$, resulting to a bound of about $850$ G\eV.

\end{abstract}

\maketitle


Investigation of the triviality of the $\lambda(\vec{\Phi}^2)^2$
theory on the lattice has resulted \cite{1,6} in an upper bound for
the Higgs mass of about $590-640$ G\eV.  A Higgs with mass of about
$640$ G\eV\  will have a width of about $130$ G\eV\  and it should be
possible to observe experimentally.  On the other hand if the Higgs is
heavier than about $820$ G\eV\  its width will be larger than about
$275$ G\eV\  (larger than a third of its mass) and it will be hard to
observe experimentally.  Recall that in a technicolor scenario the
Higgs is a very wide enhancement, analogous to the
``$\sigma$-particle'' in QCD, centered somewhere in the range 
$1$ T\eV -- $2$ T\eV.  It is therefore very important to ask 
if a lattice action can be found that will produce a heavier Higgs.

The Scalar Sector of the Minimal Standard Model is an effective theory
that describes the physics for energies less than some cutoff
$\Lambda$.  The leading cutoff effects (order $\Lambda ^{-2}$) can be
parametrized by adding to the $\lambda(\vec{\Phi}^2)^2$ theory two
bare dimension six operators with freely adjustable coefficients
\cite{2}.  This action has a total of four freely adjustable
parameters. To leading order in large $N$ we have found that the
maximum $m_H$ is obtained at $\lambda\rightarrow\infty$. In that limit
the model becomes nonlinear ($\vec{\Phi}^2=\rm{const.}$) and the
effect of the bare dimension six operators disappears (they contain
the field as $\vec{\Phi}^2$). In the nonlinear model the field
$\vec{\Phi}$ is dimensionless, and therefore the power counting
reduces to derivative counting. To parametrize the leading
$\Lambda^{-2}$ cutoff effects we now have to add two four-derivative
terms to the nonlinear action \cite{3}. This action has three freely
adjustable parameters (the fourth parameter can be thought of as being
$\lambda$ now fixed to infinity). We can then search over this three
parameter space for the maximum Higgs mass, $m_H$ under the
requirement that the observable cutoff effects are less than say
$5\%-10\%$.  This search is done first in the large $N$ \cite{4}
approximation and then using the knowledge gained it is done for the
physical $N=4$ model numerically \cite{5}.  In these proceedings we
will only present preliminary numerical results.

The action is:
\begin{eqnarray}
S&=&\int_x \ \Biggl[ {1\over 2} \vec{\Phi_c} g(-\partial^2) \vec{\Phi_c} -
{b_1 \over {2N}} 
\bigl( \partial_\mu \vec{\Phi_c} \cdot \partial_\mu\vec{\Phi_c} \bigr)^2
\nonumber \\
&& -{b_2 \over {2N}} 
\bigl(\partial_\mu \vec{\Phi_c} \cdot  \partial_\nu \vec{\Phi_c} 
-{\delta_{\mu \nu} \over 4} 
\partial_\sigma \vec{\Phi_c} \cdot \partial_\sigma \vec{\Phi_c} \bigr)^2
\Biggr]
\end{eqnarray}
where $\vec{\Phi_c} g(-\partial^2) \vec{\Phi_c}$ is a regularized
kinetic energy term, $\vec{\Phi_c}^2=N \beta$, and the partition
function is $Z=\int[d\vec{\Phi_c}]e^{-S}$. We have investigated this
action to leading order at large $N$ with Pauli Villars regularization
and have found that the phase diagram does not depend on $b_2$ (for
classical fields the term is not Lorentz invariant), and also the
ratio of the Higgs mass to the weak scale $f_\pi$ ($f_\pi=246$ G\eV)
does not depend on $b_2$ to leading order in $m_R^2$, where $m_R$ is
the renormalized mass. Under these approximations we can set $b_2=0$
and we are left with a two parameter space to search. This we have
done, but since these are the proceedings of a lattice conference we
now turn our attention to the lattice regularization.

The lattice best suited for the study of leading cutoff effects
($\Lambda^{-2}$) is the $F_4$ lattice since it does not introduce
Euclidean $O(4)$ violations by lattice artifacts to that order
\cite{2} (as the more commonly used hypercubic lattice does).
Denoting sites on $F_4$ by $x, x',x''$, and links by $<x,x'>,l,l'$,
the action is:
\begin{eqnarray}
S&=&-2N\beta_0\sum_{<x,x'>}\vec{\Phi}(x)\cdot\vec{\Phi}(x') \nonumber \\
&&-N\beta_1\sum_{<x,x'>} [\vec{\Phi}(x)\cdot\vec{\Phi}(x')]^2 \nonumber \\
&&-N{{\beta_2}\over{48}} 
\sum_x\Biggl[\sum_{{l\cap x \ne
\emptyset}\atop {l=<x,x'>} } \vec{\Phi(x})\cdot\vec{\Phi}(x')\Biggr]^2 .
\end{eqnarray}
Here the field is constrained by ${\vec{\Phi}}^2 (x) =1$. To obtain
the action in eq. $(1)$ above, the field has to be rescaled
$\vec{\Phi_c}=\sqrt {6N(\beta_0 +\beta_1 +\beta_2)}\vec{\Phi}$ (we only consider
the region $\beta_0 +\beta_1 +\beta_2 > 0$).  From the relation of the
parameters with those of action $(1)$ we find that the region of
interest $b_2=0$ corresponds to $\beta_1=0$. We then are left to
search the two parameter space $\beta_0$, $\beta_2$. We do that using
the large $N$ saddle point approximation and what follows from now on
is done in that approximation unless otherwise noted.

The phase diagram is given in fig. 1. The second order line terminates
at a tricritical point (TCP) where a first order line begins.  
The phase diagram obtained numerically for the physical $N=4$ looks very
similar to the one in fig. 1 and it has an approximately straight
second order line \cite{5}.

\begin{figure}
\epsfxsize=\columnwidth
\epsffile{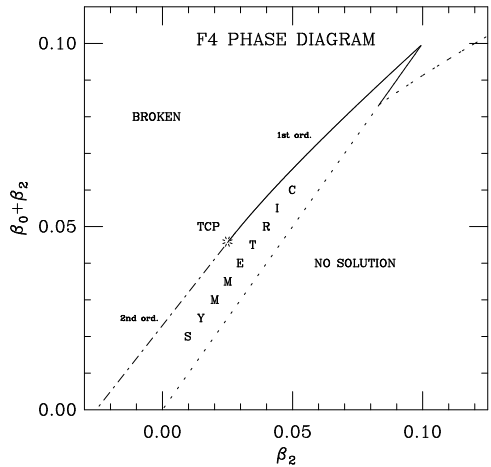}
\caption{ Phase diagram from large $N$ calculation}
\end{figure}

The renormalized mass (defined from the\break smallest positive zero
of the real part of the determinand of the matrix that appears in the
quadratic term at the end of the calculation) is a good approximation
to the Higgs mass in the perturbative regime ($f_\pi^2\sim p^2\sim
m_R^2 << 1$). We find to leading order in $m_R^2$
\begin{eqnarray}
m_R=C(\beta_2)\  exp\bigl[-16 \pi^2 {f_\pi^2 \over {N m_R^2}}\bigr]
\end{eqnarray}
with
\begin{eqnarray}
C(\beta_2)=exp\Biggl[ 8\pi^2c_1-{16\pi^2r_0^2 \over 1-{4r_0^2\over
3\beta_2}}\Biggr]
\end{eqnarray}
where $m_R$ is in lattice units, $r_0$ is a constant equal to the
momentum integral of the inverse kinetic energy term and $c_1$ is a
constant that comes from the ``bubble'' integral. Both constants have
been calculated in \cite{2}.  By demanding $\beta_2$ to be able to
stay arbitrarily close to the critical surface and at the same time
keep $\beta_0+\beta_2$ positive we find that $\beta_2$ has to be
larger than $-{4\over 3}r_0^2$ which is the point where the second
order line cuts the $x$ axis in fig.1.  As a result we find that
\begin{eqnarray}
{C(0)\over C(-{4\over 3}r_0^2)}=exp\bigl[ 8\pi^2r_0^2\bigr]=4.521
\end{eqnarray}
To check the sensitivity of keeping only leading order in $m_R^2$ we
included the next order (still neglecting $\beta_1$). The effect was
less than a $1\%-2\%$ correction on $m_R/f_\pi$ even for $m_R$ close
to $1$.  Of course one does not expect the large $N$ calculation to
give good results for $N=4$. We are however only interested in
relative changes. To that end, as a check, we compare $m_R/f_\pi$
from the $N=4$ numerical work of \cite{6} with our large $N$ result
for the case $\beta_1=\beta_2=0$.  Their relative difference is about
$25\%$ but it remains basically constant (within errors) for $m_R$ as
large as $1$.  If we restrict the cutoff effects to the differential
cross section of $\pi-\pi$ scattering at right angles with energies up
to $3m_R$ to be less than $10\%$, then we have calculated that $m_R$
must stay less than about $\sqrt{2}/2$ for the range of interest
$-{4\over 3}r_0^2 <
\beta_2 < 0$.  In fig. 2 we plot $m_R/f_\pi$ vs. $m_R$ (in lattice
units). At $m_R=\sqrt{2}/2$ the percent increase of $m_R/f_\pi$ from
$\beta_2=0$ to $\beta_2=-{4\over 3}r_0^2$ is about $30\%$ which brings
the Higgs mass bound to $\approx 850$ G\eV.

\begin{figure}
\epsfxsize=\columnwidth
\epsffile{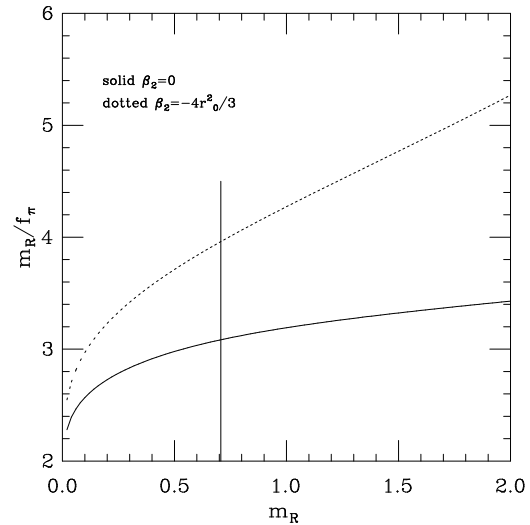}
\caption{$m_R/f_\pi$ vs. $m_R$ (in lattice units).}
\end{figure}

Preliminary numerical work for the physical case, $N=4$, \cite{5}
shows that at $m_R$ around $0.55$ the corresponding increase is about
$23\%$ within errors (from fig. 2 the large $N$ calculation gives an
increase of about $27\%$).

In conclusion, large $N$ calculations suggest that the cutoff
dependence of the effective theory (Scalar Sector of Minimal Standard
Model) can increase the triviality bound by $\approx 30 \%$. A Higgs
of that mass will have a width of $\approx 300$ G\eV\ which is more than
a third of its mass. If it turns out that this conclusion holds also
for $N=4$, as our preliminary results indicate, it will be hard to
observe the Higgs experimentally, should nature
choose to saturate the bound. 

\end{document}